# The collision frequencies in the plasmas with the power-law *q*-distributions in nonextensive statistics

Yue Wang and Jiulin Du

*Department of Physics, School of Science, Tianjin University, Tianjin 300072, China*

**Abstract** We study the collision frequencies of particles in the weakly and highly ionized plasmas with the power-law *q*-distributions in nonextensive statistics. We derive the average collision frequencies of neutral-neutral particle, electron-neutral particle, ion-neutral particle, electron- electron, ion-ion and electron-ion, respectively, in the *q*-distributed plasmas. We show that the average collision frequencies depend strongly on the *q*-parameter in a complex form and thus their properties are significantly different from that in Maxwell-distributed plasmas. These new average collision frequencies are important for us to study accurately the transport property in the complex plasmas with non-Maxwell/power-law velocity distributions.

**Key words**: Collision frequency; Power-law velocity *q*-distributions; Weakly and highly ionized plasmas; Nonextensive statistics

## 1. Introduction

Collisions of particles are common in plasmas, which contain two types: the close collisions between pairs of neutral particles or between charged particles and neutral particles, and the Coulomb collisions between pairs of charged particles. The close collisions can describe instantaneous interactions of particles at a distance comparable to the size of the particles in weakly ionized plasmas [1]. The Coulomb collisions can describe scatterings of charged particles within a distance of the Debye length [2]. The two types of collisions both can be divided into the elastic and inelastic collisions. Specifically, the elastic collisions require two particles to retain their identities and their energy states after the collisions. The inelastic collisions can lead to the ionization or the excitation [3].

The close collisions are considered for the weakly ionized plasmas with degree of ionization much less than 1 per cent, but the Coulomb collisions are considered for the highly ionized plasmas with a degree of ionization larger than 1 per cent [1]. The weakly ionized gases are of practical applications which exist in high-pressure arcs, ionospheric plasmas, process plasmas and most low-current gas discharges [3] etc. The concept of the collision cross section in weakly ionized plasmas is applied to the short-range close collisions between pairs of neutral particles or between charged particles and neutral particles, and the corresponding average collision frequencies produced by the close collisions have existed for the Maxwell-distributed plasma [4]. Highly ionized plasmas are common in universe, such as plasmas in the stellar interiors [5], plasmas in the corona of the Sun's atmosphere [6], and plasmas in vast regions of interstellar space around the hot stars of early spectral type [7] etc.

The collision frequency is an average number of collisions undergone by each particle per unit time, and so the average collision frequency in a stationary state depends on the velocity distribution functions of the colliding particles [4]. In inhomogeneous plasmas, the collisions of particles affect the transport processes and so the transport coefficients can depend significantly on formula of the average collision frequency [8-16]. The collision frequency in highly ionized plasmas is a function of the relative velocity between pairs of colliding particles [17], but that in weakly ionized plasmas is often assumed to be a constant in the Krook collision model [18]. Recently, for the weakly ionized plasmas with power-law velocity distributions, the Krook collision model was extended in the framework of nonextensive statistics and the various fluxes, such as the particle fluxes, the momentum fluxes and the heat fluxes, and the corresponding transport coefficients were studied, which of course depend on the average collision frequencies [9-14]. The collision frequencies of a test particle in highly ionized plasma are applied to understand the particle losses in a mirror machine, the energy transfer of high-energy alphas to the background plasma and the phenomenon of runaway electrons [2]. The slowing down time and collisional relaxation rate based on the Fokker-Planck collisions with Rosenbluth potential were derived for the $\kappa$-distributed and highly ionized plasmas [19,20]. The Fokker-Planck equation provides a general formula to treat the changes that result from a succession of collision in the Coulomb long-range interactions. Further, the concept of the total multiple-small-angle-collision Coulomb cross section can be applied to the long-range elastic Coulomb collisions between pairs of charged particles to demonstrate the physical process, where the corresponding average collision frequencies of charged particles due to the Coulomb collisions were derived for the Maxwell-distributed plasmas [3].

Here we focus on the average collision frequencies in the nonequilibrium complex plasmas with non-Maxwellian and/or power-law velocity distributions which exist widely in astrophysical and space plasmas [21-30]. The non-Maxwellian and/or power-law distributed complex plasmas have aroused wide interest and have generally been studied under the framework of new statistical mechanics [31]. Recently, the average collision frequency between the electrons and neutral particles was derived for the weakly ionized plasma with the velocity $q$-distribution [32], but that of electron-ion collision for a kappa-distribution was estimated very approximately [33]. But for more complex and important situations both in the weakly and highly ionized plasmas, such as the collisions between electrons, ions and neutral particles, remain to be studied. In this work, based on the different collision models in weakly and highly ionized plasmas respectively, we study the average collision frequencies of all kinds of particles in the complex plasmas.

In nonextensive statistics, the power-law velocity $q$-distribution function can be expressed [34,35] as

$$f_{q,\alpha}(\mathbf{v}) = N_\alpha B_{q,\alpha} \left(\frac{m_\alpha}{2\pi k_B T_\alpha}\right)^{\frac{3}{2}} \left[1 - (1-q_\alpha)\frac{m_\alpha \mathbf{v}^2}{2k_B T_\alpha}\right]^{\frac{1}{1-q_\alpha}}, \tag{1}$$

where $N_\alpha$ is number density, $m_\alpha$ is mass, $T_\alpha$ is temperature of the $\alpha$th particle, $k_B$ is the Boltzmann constant, $q_\alpha$ is the nonextensive parameter which characterizes the nonextensive degree of the system, and $B_{q,\alpha}$ is the $q$-dependent normalized constant given by

$$B_{q,\alpha} = \begin{cases} (1-q_\alpha)^{\frac{1}{2}}(3-q_\alpha)(5-3q_\alpha)\dfrac{\Gamma\left(\frac{1}{1-q_\alpha}+\frac{1}{2}\right)}{4\Gamma\left(\frac{1}{1-q_\alpha}\right)}, & \text{for } 0 < q_\alpha < 1. \\[1em] (q_\alpha-1)^{\frac{3}{2}}\dfrac{\Gamma\left(\frac{1}{q_\alpha-1}\right)}{\Gamma\left(\frac{1}{q_\alpha-1}-\frac{3}{2}\right)}, & \text{for } 1 < q_\alpha < \frac{5}{3}. \end{cases}$$

The nonextensive parameter $q \neq 1$ as well as the velocity $q$-distribution (1) of complex plasmas were given clear physical meanings in the nonextensive kinetics, which describe a nonequilibrium stationary-state in the complex systems with the external fields such as the electromagnetic interactions and/or the self-gravitating interactions [36-39]. And now we have known that the velocity $q$-distribution (1) is exactly equivalent to the $\kappa$-distribution observed in astrophysical and space plasmas only if we make the parameter transformation [39],

$$2T = (7-5q)\tilde{T} \text{ and } (q-1)^{-1} = \kappa + 1, \qquad (2)$$

where $\tilde{T}$ is the temperature in the $\kappa$-distribution, which is, of course, the physical temperature of the plasma, and it is also the physical temperature of the velocity $q$-distribution in nonextensive statistics. Based on this parameter transformation, the $\kappa$-distributed complex plasmas can be well studied under the framework of nonextensive statistics.

The paper is organized as follows. In Section 2, we study the average collision frequencies of particles in the weakly ionized plasma with the $q$-distribution, where the "hard-sphere" collision model is applied. In Section 3, we study the average collision frequencies of the charged particles in the highly ionized plasma with the $q$-distribution, where the Coulomb collision model is applied. In Section 4, we make numerical analyses. Finally in Section 5, we give the conclusion.

## 2. The average collision frequencies of particles in the weakly ionized plasma

In the weakly ionized plasma, the number of neutral particles is much larger than that of charged particles, the collision probability between pairs of charged particles is very small, and so the collisions of neutral-neutral particles, electron-neutral particles and ion-neutral particles dominate. In this situation, the electron, ion, or neutral particle is deflected by colliding with the neutral particle, and if they are not surrounded with the external fields of forces, only the strong short-range interactions take place by the close encounter, thus the simplest rigid, smooth and elastic hard-sphere collision model can be used [4,18]. Generally, when two particles collide, the probability of the third particle being close enough to have any effect is very small, so only the hard-sphere binary collision model is considered [1].

The effective collision frequency is defined as the average value of the product of the collision cross section and the relative velocity between the incident particle and the target particle [18]. In the hard-sphere binary collision model of the weakly ionized plasmas, the collision number between two particles denoted by $\alpha$ and $\beta$ in per unit volume and unit time is given [4] by

$$\nu_{\alpha\beta} = \iiint f_{\alpha 1}(\mathbf{v}_1) f_{\beta 2}(\mathbf{v}_2) |\mathbf{v}_1 - \mathbf{v}_2| \left(\frac{\sigma_{12}}{2}\right)^2 \sin\chi \, d\chi \, d\varepsilon \, d\mathbf{v}_1 d\mathbf{v}_2, \qquad (3)$$

where $\mathbf{v}_1$ is the velocity of the $\alpha$-particle with the velocity distribution function $f_{\alpha 1}(\mathbf{v}_1)$, $\mathbf{v}_2$ is the velocity of the $\beta$-particle with the velocity distribution function $f_{\beta 2}(\mathbf{v}_2)$, $\sigma_{12}$ is the center distance ( a half of the sum of diameters ) between the two tangent particles, and $|\mathbf{v}_1-\mathbf{v}_2|$ is the absolute value of the relative velocity vector between the velocity vectors $\mathbf{v}_1$ and $\mathbf{v}_2$, $\chi$ and $\varepsilon$ are the polar angles describing the orientation about an axis parallel to the direction of the vector $\mathbf{v}_1-\mathbf{v}_2$. For the plasma under consideration at present, the velocity distribution functions are both the power-law velocity $q$-distribution functions.

In the previous paper for the collision frequency of electron-neutral particle [32], the velocity of neutral particles was usually considered zero as compared with that of electrons, in this case the above Eq.(3) can be naturally reduced to the equation for the definition of average collision frequency in Ref.[32].

### 2.1 The average collision frequency of neutral-neutral particles

When the collisions take place between the same kinds of neutral particles in the plasma, the collision number Eq.(3) for the neutral-neutral particles in per unit volume and unit time is written as

$$\nu_{nn} = \frac{1}{2} \iiint f_{n1}(\mathbf{v}_1) f_{n2}(\mathbf{v}_2) |\mathbf{v}_1 - \mathbf{v}_2| \left(\frac{\sigma_{12}}{2}\right)^2 \sin\chi\, d\chi\, d\varepsilon\, d\mathbf{v}_1 d\mathbf{v}_2, \tag{4}$$

where each collision between the *n*-particle is counted twice, so the coefficient 1/2 appears in the expression. We substitute the *q*-distribution function (1) into Eq.(4) and integrate Eq.(4) for $\chi$ from 0 to $\pi$ and for $\varepsilon$ from 0 to $2\pi$, and we consider $|\mathbf{v}_1 - \mathbf{v}_2| = (v_1^2 + v_2^2 - 2v_1 v_2 \cos\theta)^{1/2}$, where $\theta$ is the angle between velocity vectors $\mathbf{v}_1$ and $\mathbf{v}_2$. Eq.(4) becomes

$$\nu_{q,nn} = \frac{N_n^2 m_n^3 B_{q,n}^2 \sigma_{12}^2}{4\pi (k_B T_n)^3} \iint F_n(v_1) F_n(v_2) v_1^2 dv_1 v_2^2 dv_2 \int_0^\pi (v_1^2 + v_2^2 - 2v_1 v_2 \cos\theta)^{\frac{1}{2}} \sin\theta d\theta \int_0^{2\pi} d\varphi, \tag{5}$$

where we have denoted that

$$F_n(v_1) = \left[1 - \frac{(1-q_n)m_n v_1^2}{2k_B T_n}\right]^{\frac{1}{1-q_n}}, \quad F_n(v_2) = \left[1 - \frac{(1-q_n)m_n v_2^2}{2k_B T_n}\right]^{\frac{1}{1-q_n}}.$$

In the spherical coordinate system, the velocity vector $\mathbf{v}_1$ is assumed along the positive direction of *z*-axis, $\theta$ is the angle between the velocity vector $\mathbf{v}_2$ and the positive direction of *z*-axis (i.e. the direction of $\mathbf{v}_1$), and $\varphi$ is the angle between the projection vector of $\mathbf{v}_2$ on the *x-y* plane and the positive direction of *x*-axis. The integral of angle $\theta$ is that

$$\int_0^\pi (v_1^2 + v_2^2 - 2v_1 v_2 \cos\theta)^{\frac{1}{2}} \sin\theta\, d\theta = \begin{cases} 2\left(v_1 + \dfrac{v_2^2}{3v_1}\right), & \text{for } v_1 > v_2. \\ 2\left(v_2 + \dfrac{v_1^2}{3v_2}\right), & \text{for } v_1 < v_2. \end{cases} \tag{6}$$

Then Eq.(5) is written as

$$\nu_{q,nn} = \frac{N_n^2 m_n^3 B_{q,n}^2 \sigma_{12}^2}{(k_B T_n)^3} \int_0^{v_{max}} dv_2 F_n(v_2) v_2^2 \left[\int_{v_2}^{v_{max}} dv_1 F_n(v_1)\left(v_1^3 + \frac{v_1 v_2^2}{3}\right) + \int_0^{v_2} dv_1 F_n(v_1) v_1^2 \left(v_2 + \frac{v_1^2}{3v_2}\right)\right], \tag{7}$$

with

$$v_{max} = \begin{cases} \sqrt{\dfrac{2k_B T_n}{m_n(1-q_n)}}, & 0 < q < 1. \\ +\infty, & 1 < q < \dfrac{5}{3}. \end{cases}$$

After the integrations in Eq.(7) are completed (see Appendix A), we have that

$$\nu_{q,nn} = \sqrt{\frac{2\pi k_B T_n}{m_n}} \frac{4 N_n^2 \sigma_{12}^2 (5-3q_n) B_{q,n}^2}{(2-q_n)(3-2q_n)} \begin{cases} \dfrac{1}{(1-q_n)^{\frac{5}{2}}} \dfrac{\Gamma\left(\frac{2}{1-q_n}+2\right)}{\Gamma\left(\frac{2}{1-q_n}+\frac{9}{2}\right)}, & 0 < q_n < 1. \\ \dfrac{1}{(q_n-1)^{\frac{5}{2}}} \dfrac{\Gamma\left(\frac{2}{q_n-1}-\frac{7}{2}\right)}{\Gamma\left(\frac{2}{q_n-1}-1\right)}, & 1 < q_n < \dfrac{3}{2}. \end{cases} \tag{8}$$

Since each collision affects two neutral particles at once [4], then the average collision frequency of neutral-neutral particles in the weakly ionized and *q*-distributed plasmas is obtained by

$$\bar{v}_{q,nn} = \frac{2v_{q,nn}}{N_n} = \sqrt{\frac{2\pi k_B T_n}{m_n}} \frac{8N_n \sigma_{12}^2 (5-3q_n) B_{q,n}^2}{(2-q_n)(3-2q_n)} \begin{cases} \frac{1}{(1-q_n)^{\frac{5}{2}}} \frac{\Gamma\left(\frac{2}{1-q_n}+2\right)}{\Gamma\left(\frac{2}{1-q_n}+\frac{9}{2}\right)}, & 0 < q_n < 1. \\ \frac{1}{(q_n-1)^{\frac{5}{2}}} \frac{\Gamma\left(\frac{2}{q_n-1}-\frac{7}{2}\right)}{\Gamma\left(\frac{2}{q_n-1}-1\right)}, & 1 < q_n < \frac{3}{2}. \end{cases} \quad (9)$$

It is obvious that the average collision frequency is strongly dependent on the nonextensive parameter of the plasma with the power-law velocity *q*-distribution and therefore significantly different from that of the plasma following a Maxwellian velocity distribution. Only when we take the limit $q_n \to 1$, Eq.(9) can recover to the average collision frequency of the particles with a Maxwellian velocity distribution [4],

$$\bar{v}_{1,nn} = 4N_n \sigma_{12}^2 \sqrt{\frac{\pi k_B T_n}{m_n}}. \quad (10)$$

In Ref.[40], Bezerra et al studied the thermal conductivity and viscosity coefficient for the nonextensive gas with the velocity *q*-distribution, where the average collision frequency (i.e., the reciprocal of average collision time $\tau$) was considered constant as usual. As an example of the applications of Eq.(9), the collision frequency constant in the work should be replaced by Eq.(9) and so the thermal conductivity and viscosity coefficient are modified respectively as

$$K_q = \frac{N_n k_B^2 T_n}{m_n \bar{v}_{q,nn}} \frac{10}{(7-5q_n)(9-7q_n)} \quad \text{and} \quad \eta_q = \frac{2N_n k_B T_n}{(7-5q_n)\bar{v}_{q,nn}}. \quad (11)$$

**2.2 The average collision frequencies of electron- and ion-neutral particles**

When the collisions take place between charged particles and the neutral particles in the plasma, the collision number Eq. (3) for the electron- and ion-neutral particles in per unit volume and unit time is written as

$$v_{\alpha n} = \iiint\int f_{\alpha 1}(\mathbf{v}_1) f_{n2}(\mathbf{v}_2) |\mathbf{v}_1 - \mathbf{v}_2| \left(\frac{\sigma_{12}}{2}\right)^2 \sin \chi \, d\chi \, d\varepsilon \, d\mathbf{v}_1 d\mathbf{v}_2, \quad (12)$$

where the subscript $\alpha = e, i$, denotes the electrons and ions respectively, and *n* denotes the neutral particles. Here we assume that all the particles have the same temperature and the same nonextensive parameter *q*.

In the same way from Eq.(4) to Eq.(7), Eq.(12) becomes

$$v_{q,\alpha n} = \frac{2N_\alpha N_n m_\alpha^{\frac{3}{2}} m_n^{\frac{3}{2}} B_q^2 \sigma_{12}^2}{(k_B T)^3} \int_0^{v_{n,\max}} dv_2 F_n(v_2) v_2^2 \left[ \int_{v_2}^{v_{\alpha,\max}} dv_1 F_\alpha(v_1) \left(v_1^3 + \frac{v_1 v_2^2}{3}\right) \right.$$
$$\left. + \int_0^{v_2} dv_1 F_\alpha(v_1) v_1^2 \left(v_2 + \frac{v_1^2}{3v_2}\right) \right] \quad (13)$$

with

$$v_{\alpha,\max} = \begin{cases} \sqrt{\frac{2k_B T}{m_\alpha (1-q)}}, & 0 < q < 1, \\ +\infty, & 1 < q < \frac{5}{3}, \end{cases} \quad \text{and} \quad v_{n,\max} = \begin{cases} \sqrt{\frac{2k_B T}{m_n (1-q)}}, & 0 < q < 1, \\ +\infty, & 1 < q < \frac{5}{3}, \end{cases}$$

where we have denoted that

$$F_\alpha(v_1) = \left[1 - \frac{(1-q)m_\alpha v_1^2}{2k_B T}\right]^{\frac{1}{1-q}}, \quad F_n(v_2) = \left[1 - \frac{(1-q)m_n v_2^2}{2k_B T}\right]^{\frac{1}{1-q}}.$$

If the mass of the neutral particle is greater than that of the charged particle in the plasma, i.e. $m_\alpha < m_n$, Eq.(13) is calculated for the case of $0 < q < 1$ (see Appendix B) as

$$v_{q,\alpha n} = C_q \left\{ (1-q)\left(\frac{m_\alpha}{m_n} - 1\right)\left[3(5-3q) + \frac{m_\alpha}{m_n}(q-3)\right] {}_2F_1\left(\frac{1}{2}, \frac{1}{q-1}; \frac{5}{2} + \frac{1}{1-q}; \frac{m_\alpha}{m_n}\right) \right.$$

$$\left. +2(3-2q)\left[(19-11q) + \frac{m_\alpha}{m_n}(1-q)\right] {}_2F_1\left(-\frac{1}{2}, \frac{1}{q-1}; \frac{5}{2} + \frac{1}{1-q}; \frac{m_\alpha}{m_n}\right) \right\}, \quad (14)$$

where ${}_2F_1(a,b;c;z)$ is a Hypergeometric function [41] and

$$C_q = \frac{B_q N_\alpha N_n \sigma_{12}^2 \sqrt{\frac{8\pi k_B T}{m_\alpha}}}{(2-q)(3-2q)(9-5q)(11-7q)}.$$

Similarly, Eq.(13) is calculated for the case of $q > 1$ (see Appendix B) as

$$v_{q,\alpha n} = C_q \left\{ (1-q)\left(\frac{m_\alpha}{m_n} - 1\right)\left[3(5-3q) + \frac{m_\alpha}{m_n}(q-3)\right] {}_2F_1\left(\frac{1}{2}, \frac{1}{q-1}; \frac{5}{2} + \frac{1}{1-q}; \frac{m_\alpha}{m_n}\right) \right.$$

$$\left. +2(3-2q)\left[(19-11q) + \frac{m_\alpha}{m_n}(1-q)\right] {}_2F_1\left(-\frac{1}{2}, \frac{1}{q-1}; \frac{5}{2} + \frac{1}{1-q}; \frac{m_\alpha}{m_n}\right) + A_{q,\alpha n}\left(\frac{m_\alpha}{m_n}\right)^{\frac{1}{q-1} - \frac{3}{2}} \right\}, \quad 1 < q < \frac{3}{2},$$

(15)

where we have used that

$$A_{q,\alpha n} = \frac{4\Gamma\left(\frac{1}{1-q} + \frac{7}{2}\right)\Gamma\left(\frac{2}{q-1} - \frac{3}{2}\right)}{3\sqrt{\pi}(7-5q)\Gamma\left(\frac{1}{q-1} - \frac{1}{2}\right)} \left\{ 2(7q-11)(q-3)(3q-5)\left(1 + \frac{m_\alpha}{m_n}\right) {}_2F_1\left(\frac{2}{q-1} - \frac{5}{2}, \frac{1}{q-1}; \frac{1}{q-1} - \frac{3}{2}; \frac{m_\alpha}{m_n}\right) \right.$$

$$+ 3(7q-11)(q-1)(5q-9)\left[1 + \left(\frac{m_\alpha}{m_n}\right)^2\right] {}_2F_1\left(\frac{2}{q-1} - \frac{3}{2}, \frac{1}{q-1}; \frac{1}{q-1} - \frac{1}{2}; \frac{m_\alpha}{m_n}\right)$$

$$\left. - 2(3q-5)(5q-7)(5q-9) {}_2F_1\left(\frac{2}{q-1} - \frac{7}{2}, \frac{1}{q-1}; \frac{1}{q-1} - \frac{5}{2}; \frac{m_\alpha}{m_n}\right) \right\}. \quad (16)$$

Then, the average collision frequency of the charged particle ($\alpha = e, i$) in the weakly ionized plasmas with the velocity $q$-distribution is made by

$$\bar{v}_{q,\alpha n} = \frac{v_{q,\alpha n}}{N_\alpha}. \quad (17)$$

It is obvious that the average collision frequencies of the electron- and ion-neutral particles are significantly affected by the nonextensive parameter $q$ in a very complex form. When we take limit $q \to 1$, the average collision frequencies (17) recover those in the plasma with a Maxwellian distribution [4],

$$\bar{v}_{1,\alpha n} = \frac{v_{1,\alpha n}}{N_\alpha} = N_n \sigma_{12}^2 \sqrt{8\pi k_B T \frac{(m_\alpha + m_n)}{m_\alpha m_n}}, \quad \alpha = e, i. \quad (18)$$

The average collision frequency of electron-neutral-particle in Eq.(17) is that

$$\bar{v}_{q,en} = \frac{C_q}{N_e}\left\{(1-q)\left(\frac{m_e}{m_n}-1\right)\left[3(5-3q)+\frac{m_e}{m_n}(q-3)\right]\,_2F_1\left(\frac{1}{2},\frac{1}{q-1};\frac{5}{2}+\frac{1}{1-q};\frac{m_e}{m_n}\right)\right.$$
$$\left.+2(3-2q)\left[(19-11q)+\frac{m_e}{m_n}(1-q)\right]\,_2F_1\left(-\frac{1}{2},\frac{1}{q-1};\frac{5}{2}+\frac{1}{1-q};\frac{m_e}{m_n}\right)\right\}$$
$$+\frac{C_q}{N_e}\begin{cases}0, & 0<q<1.\\ A_{q,en}\left(\frac{m_e}{m_n}\right)^{\frac{1}{q-1}-\frac{3}{2}}, & 1<q<\frac{3}{2}.\end{cases} \tag{19}$$

If we take the limit $m_e/m_n \to 0$, all the Hypergeometric functions become unity and Eq.(19) becomes the same as the previous result in Ref.[32], namely,

$$\bar{v}_{q,en} = N_n\pi\sigma_{12}^2\sqrt{\frac{8k_BT}{\pi m_e}}\begin{cases}\sqrt{\frac{1}{1-q}}\frac{\Gamma\left(\frac{1}{1-q}+\frac{5}{2}\right)}{\Gamma\left(\frac{1}{1-q}+3\right)}, & 0<q<1.\\ \sqrt{\frac{1}{q-1}}\frac{\Gamma\left(\frac{1}{q-1}-2\right)}{\Gamma\left(\frac{1}{q-1}-\frac{3}{2}\right)}, & 1<q<\frac{3}{2}.\end{cases} \tag{20}$$

where the symbol $\sigma^2_{12}$ is equivalent to the symbol $r^2_n$ in Ref.[32].

In the previous works on the transport coefficients in weakly ionized plasma with the q-distributions, such as the viscosity [10,11], the diffusion coefficient and thermal conductivity [9,12-14] etc., the average collision frequency was generally considered a constant. As examples of the applications of Eq.(17), the collision frequency constants in these works should be replaced by Eq.(17) for different cases, and then the effects of the *q*-parameter on these transport coefficients will be modified.

**3. The average collision frequencies of charged particles in the highly ionized plasma**

In the highly ionized plasma, the Coulomb collisions between charged particles become significant, which represent a basic irreducible minimum limit for transport losses in the fusion program, and thus electron-electron collisions, ion-ion collisions and electron-ion collisions become dominant [2]. Because the distant electrostatic interactions produced by the Coulomb forces between charged particles can affect each other's motion, they are not the real collisions with each other directly. The effective radius of the electrostatic interactions at ordinary temperatures is about as much as hundred times that of the collisions between charged particles and neutral particles [18]. When the incident charged particle is deflected by Coulomb collisions of target charged particles, small-angle scatterings produce a superposed 90-degree net deflection of the incident particle long before the deflection of a single large-angle close collision, so the long-range Coulomb collision model can be used in the highly ionized plasmas [1,3]. Further, each small-angle deflection represents a small perturbation about the trajectory, and thus the combined result can be treated as the superposition of two-body interactions [2].

The Coulomb collisions in the highly ionized plasma contain two qualitatively different types of transport: velocity space transport and physical space transport, respectively. In the velocity space transport, the momentum exchange and the energy exchange correspond to the different

Coulomb cross sections, respectively [2]. In the momentum exchange between pairs of charged particles, the momentum relaxation of the incident particle represents the total loss of directed momentum of the incident particle. When the incident charged particle passes through the background of field charged particles, the momentum relaxation collision frequency can be calculated by the test particle method [1]. The cross section of the test particle is the Coulomb cross section, which is a function of the relative velocity between the test particle and the target particle. In fact, the test particle in highly ionized plasmas is shielded from the electric field of all target particles far away than a Debye length [2]. For the momentum relaxation of the test particle $\alpha$ in highly ionized plasmas, the momentum relaxation collision frequency in the background of field charged particles $\beta$ can be expressed [3,17,42-43] as

$$\nu_{\alpha\beta} = \frac{N_\beta Z_\alpha^2 Z_\beta^2 e^4 \ln\Lambda}{4\pi\varepsilon_0^2 m_{\alpha\beta}^2 v_{\alpha\beta}^3}, \qquad (21)$$

where the reduced mass is $m_{\alpha\beta}=m_\alpha m_\beta/(m_\alpha+m_\beta)$, the absolute value of relative velocity vector is $v_{\alpha\beta} = |\mathbf{v}_\alpha - \mathbf{v}_\beta|$, the charge of an $\alpha$-particle is $Z_\alpha e$, the charge of a $\beta$-particle is $Z_\beta e$, $\varepsilon_0$ is the vacuum dielectric constant, $N_\beta$ is the number density of the $\beta$th particle, and $\ln\Lambda$ is the Coulomb scattering factor determined by the Debye length.

In Eq.(21), the collision frequency $\nu_{\alpha\beta}$ is a function of the relative velocity $v_{\alpha\beta}$ between the interacting particles, thus the average collision frequency will depend on the velocity distributions of the charged particles in the plasma under consideration. Namely, the average collision frequency of the $\alpha$th particle is

$$\bar{\nu}_{\alpha\beta} = \frac{N_\beta Z_\alpha^2 Z_\beta^2 e^4 \ln\Lambda}{4\pi\varepsilon_0^2 m_{\alpha\beta}^2}\langle v_{\alpha\beta}^{-3}\rangle, \qquad (22)$$

When the plasma is in the thermal equilibrium state and so has a Maxwell velocity distribution, the average collision frequency is given [3] by

$$\bar{\nu}_{1,\alpha\alpha} = \frac{N_\alpha Z_\alpha^4 e^4 \ln\Lambda}{12\varepsilon_0^2 m_\alpha^{1/2}(\pi k_B T_\alpha)^{3/2}}, \quad \alpha=e,i, \qquad (23)$$

for the electron-electron and ion-ion collisions, and

$$\bar{\nu}_{1,ei} = \frac{\sqrt{2} N_i Z_i^2 e^4 \ln\Lambda}{12\varepsilon_0^2 m_e^{1/2}(\pi k_B T_e)^{3/2}}, \qquad (24)$$

for the electron-ion collisions. But if the plasma is not in the thermal equilibrium state and has a non-Maxwell velocity distribution, the average collision frequency needs to be recalculated on the basis of the velocity distribution functions of particles.

**3. 1 The average collision frequency of electron-ion**

According to Eq.(22), the average collision frequency of electron-ion in the plasma with the power-law velocity $q$-distribution is calculated [44] by

$$\bar{\nu}_{ei} = \frac{N_i Z_i^2 e^4 \ln\Lambda}{4\pi\varepsilon_0^2 m_{ei}^2}\langle v_{ei}^{-3}\rangle_q. \qquad (25)$$

Just as in the Maxwell-distributed plasma, on the right-hand side of Eq.(25), the integral calculation of the relative velocity for the $q$-distribution is still divergent. To overcome this problem, usually the average collision frequency is derived by means of calculating the interaction force between the charged particles. Then when (light) electrons are striking (heavy) ions, the interaction force exerted by ions on the electrons in the center-of-mass frame can be given [3] as

$$\mathbf{F}_{ei} = -N_e m_{ei} \langle \nu_{ei} \mathbf{v}_{ei} \rangle_q, \tag{26}$$

where the average value is defined as [45]

$$\langle \nu_{ei} \mathbf{v}_{ei} \rangle_q = \frac{\int \nu_{ei} \mathbf{v}_{ei} f_{q,e} d\mathbf{v}}{\int f_{q,e} d\mathbf{v}}, \tag{27}$$

the reduced mass is $m_{ei}=m_e m_i/(m_e+m_i)\approx m_e$, and the relative velocity vector is $\mathbf{v}_{ei}=\mathbf{v}_e-\mathbf{v}_i\approx \mathbf{v}_e$. For present purposes, we suppose the drifting electrons have a non-zero mean velocity vector $\mathbf{u}$, accordingly, the velocity $q$-distribution function is written as

$$f_{q,e}(\mathbf{v}_e) = N_e B_{q,e} \left(\frac{m_e}{2\pi k_B T_e}\right)^{\frac{3}{2}} \left[1-(1-q)\frac{m_e(\mathbf{v}_e-\mathbf{u})^2}{2k_B T_e}\right]^{\frac{1}{1-q}}. \tag{28}$$

Since the non-zero mean velocity vector $\mathbf{u}$ has no effects on the average collision frequency, for simplification [3], it is assumed to be $\mathbf{u}=(0, 0, u_z)$ and $u_z$ is very small as compared with the thermal velocity, namely, $u_z<<(k_B T_e/m_e)^{1/2}$. Eq.(28) can be expanded up to the first order of $\mathbf{u}$ as

$$f_{q,e}(\mathbf{v}_e) \approx N_e B_{q,e} \left(\frac{m_e}{2\pi k_B T_e}\right)^{\frac{3}{2}} \left\{1+\frac{m_e \mathbf{v}_e \cdot \mathbf{u}}{k_B T_e}\left[1-(1-q)\frac{m_e v_e^2}{2k_B T_e}\right]^{-1}\right\}\left[1-(1-q)\frac{m_e v_e^2}{2k_B T_e}\right]^{\frac{1}{1-q}}. \tag{29}$$

On the basis of Eq.(21), the collision frequency is written as

$$\nu_{ei} = \frac{N_i Z_i^2 e^4 \ln\Lambda}{4\pi\varepsilon_0^2 m_{ei}^2 v_{ei}^3} \approx \frac{N_i Z_i^2 e^4 \ln\Lambda}{4\pi\varepsilon_0^2 m_e^2 v_e^3}. \tag{30}$$

Substituting Eqs.(29) and (30) into Eq.(26), we derive (see Appendix C) that

$$F_{ei,z} = -\frac{\sqrt{2}m_e^{\frac{1}{2}} u_z N_e N_i Z_i^2 e^4 \ln\Lambda}{12\varepsilon_0^2 (\pi k_B T_e)^{\frac{3}{2}}} \begin{cases} \dfrac{(1-q)^{\frac{1}{2}}(3-q)(5-3q)\Gamma\left(\frac{1}{1-q}+\frac{1}{2}\right)}{4\Gamma\left(\frac{1}{1-q}\right)}, 0<q<1. \\ \dfrac{(q-1)^{\frac{3}{2}}\Gamma\left(\frac{1}{q-1}\right)}{\Gamma\left(\frac{1}{q-1}-\frac{3}{2}\right)}, 1<q<\frac{5}{3}. \end{cases} \tag{31}$$

On the other hand [3], Eq.(26) is equal to

$$F_{ei,z} = -N_e m_e \langle \nu_{ei} \rangle_q u_z. \tag{32}$$

Comparing Eq.(31) with Eq.(32), the average collision frequency of the electron-ion collisions is

$$\bar{\nu}_{q,ei} = \langle \nu_{ei} \rangle_q = \frac{\sqrt{2} N_i Z_i^2 e^4 \ln\Lambda}{12\varepsilon_0^2 m_e^{\frac{1}{2}} (\pi k_B T_e)^{\frac{3}{2}}} \begin{cases} \dfrac{(1-q)^{\frac{1}{2}}(3-q)(5-3q)\Gamma\left(\frac{1}{1-q}+\frac{1}{2}\right)}{4\Gamma\left(\frac{1}{1-q}\right)}, 0<q<1. \\ \dfrac{(q-1)^{\frac{3}{2}}\Gamma\left(\frac{1}{q-1}\right)}{\Gamma\left(\frac{1}{q-1}-\frac{3}{2}\right)}, 1<q<\frac{5}{3}. \end{cases} \tag{33}$$

It is clear that the average collision frequency is strongly dependent on the $q$-parameter (the power-law index) $q\neq 1$ in a complex form. When we take limit $q\rightarrow 1$, Eq.(33) becomes Eq.(24) in the plasma with a Maxwellian velocity distribution.

### 3. 2 The average collision frequencies of electron-electron and ion-ion

According to Eq.(22), the average collision frequencies of electron-electron and ion-ion collisions in the plasma with the power-law velocity $q$-distribution are calculated by

$$\bar{\nu}_{q,\alpha\alpha} = \langle \nu_{\alpha\alpha} \rangle_q = \frac{N_\alpha Z_\alpha^4 e^4 \ln \Lambda}{4\pi\varepsilon_0^2 m_{\alpha\alpha}^2} \langle v_{\alpha\alpha}^{-3} \rangle_q, \alpha = e, i. \quad (34)$$

For the same reason as that in Eq.(25), the average collision frequencies are derived by means of calculating the interaction force between the same kind of charged particles. The interaction force in the center-of-mass frame can be given [3] as

$$\mathbf{F}_{\alpha\alpha} = -N_\alpha m_{\alpha\alpha} \langle \nu_{\alpha\alpha} \mathbf{v}_{\alpha\alpha} \rangle_q, \quad (35)$$

where, on the basis of Eq.(21), the collision frequency is written as

$$\nu_{\alpha\alpha} = \frac{N_\alpha Z_\alpha^4 e^4 \ln \Lambda}{4\pi\varepsilon_0^2 m_{\alpha\alpha}^2 v_{\alpha\alpha}^3} = \frac{N_\alpha Z_\alpha^4 e^4 \ln \Lambda}{\pi\varepsilon_0^2 m_\alpha^2 |\mathbf{v}_{\alpha,1} - \mathbf{v}_{\alpha,2}|^3}. \quad (36)$$

The power-law velocity $q$-distribution function of the pair of charged particles is taken as

$$f_q(\mathbf{v}_{\alpha\alpha}) = N_\alpha B_{q,\alpha} \left(\frac{m_{\alpha\alpha}}{2\pi k_B T_\alpha}\right)^{\frac{3}{2}} \left[1 - (1-q)\frac{m_{\alpha\alpha}(\mathbf{v}_{\alpha\alpha} - \mathbf{u})^2}{2k_B T_\alpha}\right]^{\frac{1}{1-q}}. \quad (37)$$

The idea and the way for the calculation of the interaction force Eq.(35) are similar to those of Eq.(26) in section 3.1. Thus, after the calculations we can derive that

$$F_{\alpha\alpha,z} = -\frac{m_\alpha^{\frac{1}{2}} u_z N_\alpha^2 Z_\alpha^4 e^4 \ln \Lambda}{12\varepsilon_0^2 (\pi k_B T_\alpha)^{\frac{3}{2}}} \begin{cases} \frac{(1-q)^{\frac{1}{2}}(3-q)(5-3q)\Gamma\left(\frac{1}{1-q}+\frac{1}{2}\right)}{4\Gamma\left(\frac{1}{1-q}\right)}, 0 < q < 1. \\ \frac{(q-1)^{\frac{3}{2}}\Gamma\left(\frac{1}{q-1}\right)}{\Gamma\left(\frac{1}{q-1}-\frac{3}{2}\right)}, 1 < q < \frac{5}{3}. \end{cases} \quad (38)$$

On the other hand, because the mean velocity vector $\mathbf{u}$ is assumed to be along $z$-axis, the interaction force [3] is

$$F_{\alpha\alpha,z} = -N_\alpha m_\alpha \langle \nu_{\alpha\alpha} \rangle_q u_z. \quad (39)$$

Comparing Eq.(38) with Eq.(39), we obtain the average collision frequencies for the electron-electron and ion-ion collisions ($\alpha=e, i$),

$$\bar{\nu}_{q,\alpha\alpha} = \langle \nu_{\alpha\alpha} \rangle_q = \frac{N_\alpha Z_\alpha^4 e^4 \ln \Lambda}{12\varepsilon_0^2 m_\alpha^{\frac{1}{2}}(\pi k_B T_\alpha)^{\frac{3}{2}}} \begin{cases} \frac{(1-q)^{\frac{1}{2}}(3-q)(5-3q)\Gamma\left(\frac{1}{1-q}+\frac{1}{2}\right)}{4\Gamma\left(\frac{1}{1-q}\right)}, 0 < q < 1. \\ \frac{(q-1)^{\frac{3}{2}}\Gamma\left(\frac{1}{q-1}\right)}{\Gamma\left(\frac{1}{q-1}-\frac{3}{2}\right)}, 1 < q < \frac{5}{3}. \end{cases} \quad (40)$$

Thus the average collision frequencies given in Eq.(40) are significantly dependent on the nonextensive parameter $q \neq 1$ (the power-law index) in a complex form. When we take limit $q \to 1$, Eq.(40) becomes Eq.(23) in the plasma with a Maxwellian velocity distribution.

**4. Numerical analyses of the average collision frequencies**

In order to show the effects more clearly of the nonextensive parameter on the average collision frequencies of the particles in the plasma with the power-law velocity $q$-distributions, now we make numerical analyses of the average collision frequencies of the neutral-neutral particle, the electron-neutral particle, the ion-neutral particle, the electron-ion, the ion-ion and the electron-electron, respectively.

For this purpose, using Eq.(9), Eq.(10), Eq.(17) and Eq.(18), we can write the relations for the weakly ionized plasma with the power-law $q$-distribution,

$$\frac{\bar{v}_{q,nn}}{\bar{v}_{1,nn}} = \frac{2\sqrt{2}(5-3q_n)B_{q,n}^2}{(2-q_n)(3-2q_n)} \begin{cases} (1-q_n)^{-\frac{5}{2}} \frac{\Gamma\left(\frac{2}{1-q_n}+2\right)}{\Gamma\left(\frac{2}{1-q_n}+\frac{9}{2}\right)}, & 0 < q_n < 1, \\ (q_n-1)^{-\frac{5}{2}} \frac{\Gamma\left(\frac{2}{q_n-1}-\frac{7}{2}\right)}{\Gamma\left(\frac{2}{q_n-1}-1\right)}, & 1 < q_n < \frac{3}{2}, \end{cases} \quad (41)$$

$$\frac{\bar{v}_{q,\alpha n}}{\bar{v}_{1,\alpha n}} = \frac{(1+m_\alpha/m_n)^{-1/2} B_q}{(2-q)(3-2q)(9-5q)(11-7q)}$$

$$\left\{(1-q)\left(\frac{m_\alpha}{m_n}-1\right)\left[3(5-3q)+\frac{m_\alpha}{m_n}(q-3)\right] {}_2F_1\left(\frac{1}{2},\frac{1}{q-1};\frac{5}{2}+\frac{1}{1-q};\frac{m_\alpha}{m_n}\right)\right.$$

$$\left. + 2(3-2q)\left[(19-11q)+\frac{m_\alpha}{m_n}(1-q)\right] {}_2F_1\left(-\frac{1}{2},\frac{1}{q-1};\frac{5}{2}+\frac{1}{1-q};\frac{m_\alpha}{m_n}\right) + \left(\frac{m_\alpha}{m_n}\right)^{\frac{1}{q-1}-\frac{3}{2}} A_{q,\alpha n}\right\}, 0 < q < \frac{3}{2},$$

(42)

which determine the roles of the nonextensive parameters in the average collision frequencies of the neutral-neutral particle, the electron-neutral particle and the ion-neutral particle, respectively. In Eq.(42), $A_{q,\alpha n}=0$ for the case of $0 < q < 1$ and $A_{q,\alpha n}$ is given by Eq.(16) for the case of $1 < q < 3/2$.

Using Eq.(33), Eq.(40), Eq.(23) and Eq.(24), we can write the relation for the highly ionized plasma with the power-law $q$-distribution, $\alpha=e, i$,

$$\frac{\bar{v}_{q,ei}}{\bar{v}_{1,ei}} = \frac{\bar{v}_{q,\alpha\alpha}}{\bar{v}_{1,\alpha\alpha}} = \begin{cases} \dfrac{(1-q)^{\frac{1}{2}}(3-q)(5-3q)\Gamma\left(\frac{1}{1-q}+\frac{1}{2}\right)}{4\Gamma\left(\frac{1}{1-q}\right)}, & 0 < q < 1, \\ \dfrac{(q-1)^{\frac{3}{2}} \Gamma\left(\frac{1}{q-1}\right)}{\Gamma\left(\frac{1}{q-1}-\frac{3}{2}\right)}, & 1 < q < \frac{5}{3}, \end{cases} \quad (43)$$

which determines the roles of the nonextensive parameters in the average collision frequencies of the electron-ion, the ion-ion and the electron-electron, respectively.

Based on Eq.(41), in Fig.1 the numerical analyses show the role of the nonextensive parameter in the average collision frequency between pairs of the neutral particles with the same mass in the weakly ionized plasma with the velocity $q$-distribution, where $v_{q,nn}/v_{1,nn}$ is the ordinate axis and the nonextensive parameter $q_n$ is the abscissa axis. It is shown that the average collision frequency of neutral-neutral particles increases monotonously with the $q$-parameter increases. In the case of $0 < q_n < 1$, the average collision frequency in the $q$-distributed plasma is slightly less than that in the Maxwell-distributed plasma, but in the case of $1 < q_n < 3/2$, the average collision frequency in the $q$-distributed plasma is more than that in the Maxwell-distributed plasma, and increases rapidly with the $q$-parameter increases.

Based on Eq.(42), in Fig.2 the numerical analyses show the roles of the $q$-parameter in the average collision frequencies of the electron- and ion-neutral particle in the weakly ionized plasma with the velocity $q$-distribution, where $v_{q,\alpha n}/v_{1,\alpha n}$ ($\alpha=e, i$) is the ordinate axis, the mass ratio $m_\alpha/m_n$ ($\alpha=e, i$) is the abscissa axis, the $q$-parameter takes five different values respectively and $q=1$ corresponds to the situation of the Maxwell-distributed plasma. It is shown that the average collision frequencies in the *q*-distributed plasma are not strongly dependent on the mass ratio of the two colliding particles, but they are strongly dependent on the *q*-parameter and increase with the *q*-parameter increases. When $0 < q < 1$, the average collision frequencies in the *q*-distributed plasma are slightly less than those in a Maxwell-distributed plasma. When $q > 1$, the average collision frequencies in the *q*-distributed plasma increase rapidly with the *q*-parameter increases and are more than those in the Maxwell-distributed plasma. When $m_\alpha/m_n \ll 1$, the curves character the average collision frequency of electron-neutral particles, but when $m_\alpha/m_n \to 1$, the curves character the average collision frequency of ion-neutral particles.

Based on Eq.(43), in Fig.3 the numerical analyses show the role of the *q*-parameter in the average collision frequencies of the electron-ion collisions, ion-ion collisions and electron-electron collisions in the highly ionized plasma with the power-law velocity *q*-distribution. It is shown that the average collision frequencies decrease monotonously with the *q*-parameter increases. In the case of $0 < q < 1$, the average collision frequencies in the *q*-distributed plasma are more than those in the Maxwell-distributed plasma, but in the case of $1 < q < 5/3$, the average collision frequencies in the *q*-distributed plasma are less than those in the Maxwell-distributed plasma.

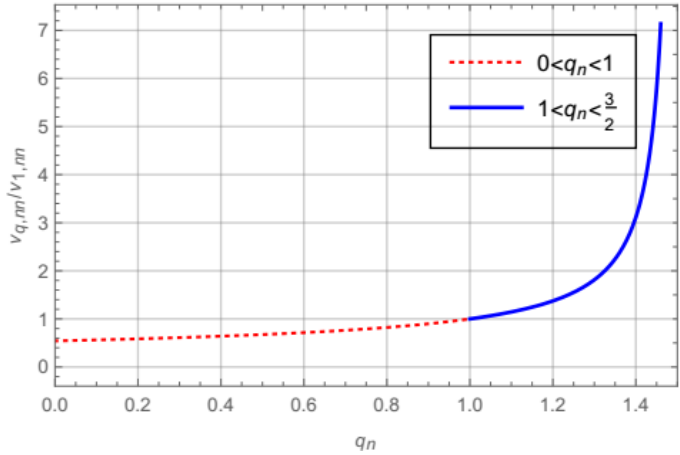

Fig. 1. The role of the nonextensive *q*-parameter in the average collision frequency $v_{q,nn}$.

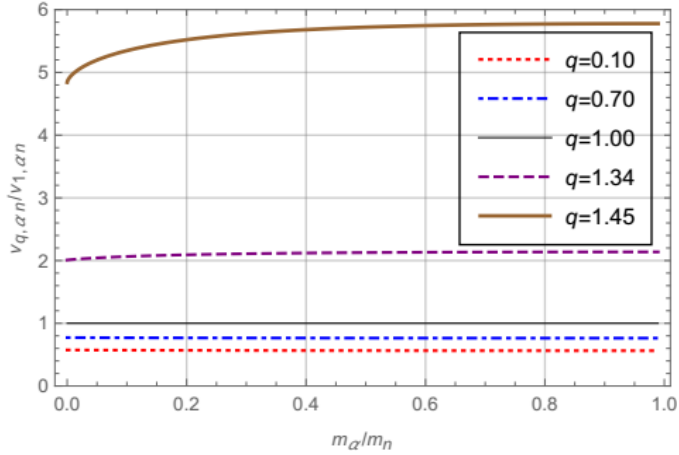

Fig. 2. The roles of the mass ratio and the nonextensive *q*-parameter in the
average collision frequency $v_{q,\alpha n}$ for $\alpha = e, i$.

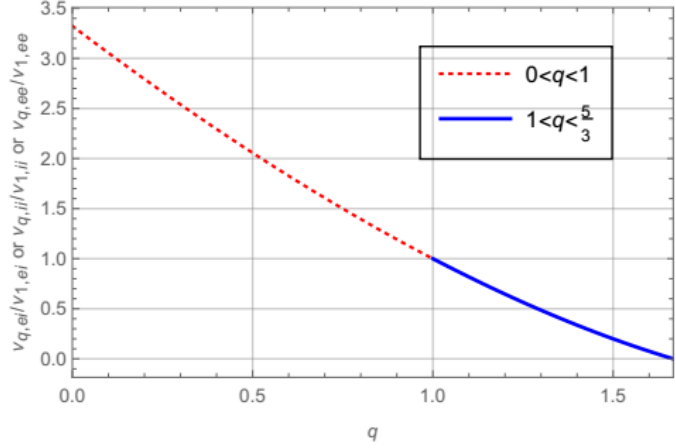

Fig. 3. The role of the nonextensive $q$-parameter in the average collision

frequency $v_{q,ei}$, $v_{q,ii}$ and $v_{q,ee}$.

In summary, we show that the nonextensive $q$-parameter $q\neq 1$ plays a significant role in the average collision frequencies of particles in the plasma with the power-law velocity $q$-distribution. But the dependence on the $q$-parameter of the average collision frequencies between two charged particles in the highly ionized plasma is quite different from those between a charged particle and a neutral particle in the weakly ionized plasma. And the average collision frequencies between the charged particle and the neutral particle in the $q$-distributed plasma are almost not affected by the mass ratio between the two colliding particles. Namely, the average collision frequency of electron-neutral particle is basically the same as that of ion-neutral particle.

## 5. Conclusion

To sum up, we have studied the collision frequencies between the particles in the plasma with the velocity $q$-distribution in nonextensive statistics. As usual, in the weakly ionized plasma, the collision model of neutral-neutral particle, electron-neutral particle and ion-neutral particle is taken the "hard-sphere" model with the short-range close collision. In the highly ionized plasma, the collision model of electron-ion, electron-electron and ion-ion is taken the Coulomb collision model with the long-range Coulomb scatterings.

We derived the average collision frequencies of neutral-neutral particle, electron-neutral particle and ion-neutral particle in the weakly ionized plasma with the velocity $q$-distribution, expressed by Eq.(9) and Eq.(17) respectively. And then we also derived the average collision frequencies of electron-ion, electron-electron and ion-ion in the highly ionized plasma with the velocity $q$-distribution, expressed by Eq.(33) and Eq.(40) respectively. We find that these new formulae of the average collision frequencies depend strongly on the nonextensive $q$-parameter $q \neq 1$ in a very complex form and thus are significantly different from those based on a Maxwellian velocity distribution. These generalized collision frequencies based on the power-law velocity $q$-distributions are important for us to study accurately the transport processes and transport coefficients in the nonequilibrium and power-law-distributed complex plasmas.

In the numerical analyses, we showed that in the weakly ionized plasma, the average collision frequencies of neutral-neutral particle, electron-neutral particle and ion-neutral particle increase with the $q$-parameter increases. Therefore, in the case of $q <1$, the average collision frequency in the $q$-distributed plasma is less than that in the Maxwell-distributed plasma, but in the case of *q* >1, the average collision frequency in the *q*-distributed plasma is more than that in the Maxwell-distributed plasma. But, in the highly ionized plasma, the average collision frequencies of electron-ion, electron-electron and ion-ion decrease monotonously with the *q*-parameter increases. Therefore, in the case of *q* <1, the average collision frequency in the *q*-distributed plasma is more than that in the Maxwell-distributed plasma, but in the case of 1< *q* <5/3, the average collision frequency in the *q*-distributed plasma is less than that in the Maxwell-distributed plasma. So the dependence on the *q*-parameter of the average collision frequencies between two charged particles in the highly ionized plasma is quite different from those between a charged particle and a neutral particle in the weakly ionized plasma.

**Acknowledgement**

This work is supported by the National Natural Science Foundation of China under Grant No. 11775156.

**Appendix A**

In Eq.(7), by using the integral transformation,

$$\int_0^{v_{max}} dv_2 F_n(v_2) v_2^2 \int_0^{v_2} dv_1 F_n(v_1) v_1^2 \left( v_2 + \frac{v_1^2}{3v_2} \right) = \int_0^{v_{max}} dv_1 F_n(v_1) v_1^2 \int_{v_1}^{v_{max}} dv_2 F_n(v_2) v_2^2 \left( v_2 + \frac{v_1^2}{3v_2} \right),$$

(A.1)

we have that

$$\nu_{q,nn} = \frac{N_n^2 m_n^3 B_{q,n}^2 \sigma_{12}^2}{(k_B T_n)^3} \left[ \int_0^{v_{max}} dv_2 F_n(v_2) v_2^2 \int_{v_2}^{v_{max}} dv_1 F_n(v_1) \left( v_1^3 + \frac{v_1 v_2^2}{3} \right) \right.$$

$$\left. + \int_0^{v_{max}} dv_1 F_n(v_1) v_1^2 \int_{v_1}^{v_{max}} dv_2 F_n(v_2) v_2^2 \left( v_2 + \frac{v_1^2}{3v_2} \right) \right]$$

$$= \frac{2 N_n^2 m_n B_{q,n}^2 \sigma_{12}^2}{(2-q_n)(3-2q_n) k_B T_n} \left[ \int_0^{v_{max}} dv_2 \left[ 1 + \frac{(9-5q_n) m_n v_2^2}{6 k_B T_n} \right] \left[ 1 - \frac{(1-q_n) m_n v_2^2}{2 k_B T_n} \right]^{\frac{2}{1-q_n}+1} v_2^2 \right.$$

$$\left. + \int_0^{v_{max}} dv_1 \left[ 1 + \frac{(9-5q_n) m_n v_1^2}{6 k_B T_n} \right] \left[ 1 - \frac{(1-q_n) m_n v_1^2}{2 k_B T_n} \right]^{\frac{2}{1-q_n}+1} v_1^2 \right]$$

$$= \frac{4 N_n^2 m_n B_{q,n}^2 \sigma_{12}^2}{(2-q_n)(3-2q_n) k_B T_n} \int_0^{v_{max}} dv_2 \left[ 1 - \frac{(1-q_n) m_n v_2^2}{2 k_B T_n} \right]^{\frac{2}{1-q_n}+1} \left[ 1 + \frac{(9-5q_n) m_n v_2^2}{6 k_B T_n} \right] v_2^2. \quad (A.2)$$

For 0< $q_n$ <1, we have

$$v_{max} = \sqrt{\frac{2 k_B T_n}{m_n (1-q_n)}},$$

and then Eq.(A.2) becomes

$$\nu_{q,nn} = \frac{4 N_n^2 m_n B_{q,n}^2 \sigma_{12}^2}{(2-q_n)(3-2q_n) k_B T_n} \int_0^{\sqrt{\frac{2 k_B T_n}{m_n(1-q_n)}}} dv_2 \left[ 1 - \frac{(1-q_n) m_n v_2^2}{2 k_B T_n} \right]^{\frac{2}{1-q_n}+1} \left[ 1 + \frac{(9-5q_n) m_n v_2^2}{6 k_B T_n} \right] v_2^2$$

$$= \sqrt{\frac{2\pi k_B T_n}{m_n}} \frac{4N_n^2 \sigma_{12}^2 B_{q,n}^2 (5-3q_n)}{(2-q_n)(3-2q_n)(1-q_n)^{\frac{5}{2}}} \frac{\Gamma\left(\frac{2}{1-q_n}+2\right)}{\Gamma\left(\frac{2}{1-q_n}+\frac{9}{2}\right)}. \quad \text{(A.3)}$$

Similarly, for $q_n > 1$, we have $v_{\max} = +\infty$. Eq.(A.2) becomes

$$v_{q,nn} = \frac{4N_n^2 m_n B_{q,n}^2 \sigma_{12}^2}{(2-q_n)(3-2q_n)k_B T_n} \int_0^{+\infty} dv_2 \left[1 - \frac{(1-q_n)m_n v_2^2}{2k_B T_n}\right]^{\frac{2}{1-q_n}+1} \left[1 + \frac{(9-5q_n)m_n v_2^2}{6k_B T_n}\right] v_2^2$$

$$= \sqrt{\frac{2\pi k_B T_n}{m_n}} \frac{4N_n^2 \sigma_{12}^2 B_{q,n}^2 (5-3q_n)}{(2-q_n)(3-2q_n)(q_n-1)^{\frac{5}{2}}} \frac{\Gamma\left(\frac{2}{q_n-1}-\frac{7}{2}\right)}{\Gamma\left(\frac{2}{q_n-1}-1\right)}, \quad 1 < q_n < \frac{3}{2}. \quad \text{(A.4)}$$

Then, substituting the expression of $B_{q,n}$ in Eq.(1) into Eqs.(A.3) and (A.4), we obtain Eq.(8).

**Appendix B**

In Eq.(13), by using the integral transformation, for $0 < q < 1$, we have

$$\int_0^{\sqrt{\frac{2k_B T}{m_n(1-q)}}} dv_2 F_n(v_2) v_2^2 \int_0^{v_2} dv_1 F_\alpha(v_1) v_1^2 \left(v_2 + \frac{v_1^2}{3v_2}\right)$$
$$= \int_0^{\sqrt{\frac{2k_B T}{m_n(1-q)}}} dv_1 F_\alpha(v_1) v_1^2 \int_{v_1}^{\sqrt{\frac{2k_B T}{m_n(1-q)}}} dv_2 F_n(v_2) v_2^2 \left(v_2 + \frac{v_1^2}{3v_2}\right), \quad \text{(B.1)}$$

so we have that

$$v_{q,\alpha n} = \frac{2N_\alpha N_n m_\alpha^{\frac{3}{2}} m_n^{\frac{3}{2}} B_q^2 \sigma_{12}^2}{(k_B T)^3} \left[\int_0^{\sqrt{\frac{2k_B T}{m_n(1-q)}}} dv_2 F_n(v_2) v_2^2 \int_{v_2}^{\sqrt{\frac{2k_B T}{m_\alpha(1-q)}}} dv_1 F_\alpha(v_1) \left(v_1^3 + \frac{v_1 v_2^2}{3}\right) \right.$$

$$\left. + \int_0^{\sqrt{\frac{2k_B T}{m_n(1-q)}}} dv_1 F_\alpha(v_1) v_1^2 \int_{v_1}^{\sqrt{\frac{2k_B T}{m_n(1-q)}}} dv_2 F_n(v_2) v_2^2 \left(v_2 + \frac{v_1^2}{3v_2}\right)\right]. \quad \text{(B.2)}$$

Then Eq.(B.2) becomes

$$v_{q,\alpha n} = \frac{2N_\alpha N_n m_\alpha^{\frac{3}{2}} m_n^{\frac{3}{2}} B_q^2 \sigma_{12}^2}{(k_B T)^3} \left[\int_0^{\sqrt{\frac{2k_B T}{m_n(1-q)}}} dv_2 F_n(v_2) \int_{v_2}^{\sqrt{\frac{2k_B T}{m_\alpha(1-q)}}} dv_1 \left[1 - \frac{(1-q)m_\alpha v_1^2}{2k_B T}\right]^{\frac{1}{1-q}} \left(v_2^2 v_1^3 + \frac{v_1 v_2^4}{3}\right)\right.$$

$$\left. + \int_0^{\sqrt{\frac{2k_B T}{m_n(1-q)}}} dv_1 F_\alpha(v_1) \int_{v_1}^{\sqrt{\frac{2k_B T}{m_n(1-q)}}} dv_2 \left[1 - \frac{(1-q)m_n v_2^2}{2k_B T}\right]^{\frac{1}{1-q}} \left(v_1^2 v_2^3 + \frac{v_2 v_1^4}{3}\right)\right]$$

$$= \frac{4N_\alpha N_n m_\alpha^{\frac{3}{2}} m_n^{\frac{3}{2}} B_q^2 \sigma_{12}^2}{(2-q)(3-2q)k_B T} \left\{\int_0^{\sqrt{\frac{2k_B T}{m_n(1-q)}}} dv_2 \frac{v_2^2}{m_\alpha^2}\left[1 - \frac{(1-q)m_n v_2^2}{2k_B T}\right]^{\frac{1}{1-q}}\left[1 + \frac{(9-5q)m_\alpha v_2^2}{6k_B T}\right]\left[1 - \frac{(1-q)m_\alpha v_2^2}{2k_B T}\right]^{\frac{1}{1-q}+1}\right.$$

$$\left. + \int_0^{\sqrt{\frac{2k_B T}{m_n(1-q)}}} dv_1 \frac{v_1^2}{m_n^2}\left[1 - \frac{(1-q)m_\alpha v_1^2}{2k_B T}\right]^{\frac{1}{1-q}}\left[1 + \frac{(9-5q)m_n v_1^2}{6k_B T}\right]\left[1 - \frac{(1-q)m_n v_1^2}{2k_B T}\right]^{\frac{1}{1-q}+1}\right\}$$

$$= \frac{4N_\alpha N_n m_\alpha^{\frac{3}{2}} m_n^{\frac{3}{2}} B_q^2 \sigma_{12}^2}{(2-q)(3-2q)k_B T} \int_0^{\sqrt{\frac{2k_B T}{m_n(1-q)}}} dv_1 g_{\alpha n}(v_1) \left\{\left[1 - \frac{(1-q)m_\alpha v_1^2}{2k_B T}\right]\left[1 - \frac{(1-q)m_n v_1^2}{2k_B T}\right]\right\}^{\frac{1}{1-q}}, \quad \text{(B.3)}$$

where
$$g_{\alpha n}(v_1) = \left(\frac{1}{m_\alpha^2} + \frac{1}{m_n^2}\right)v_1^2 + \left(\frac{1}{m_\alpha} + \frac{1}{m_n}\right)\left(1 - \frac{q}{3}\right)\frac{v_1^4}{k_B T} - \frac{(9-5q)(1-q)}{6(k_B T)^2}v_1^6.$$

Further, Eq.(B.3) becomes

$$\nu_{q,\alpha n} = C_q \left\{(1-q)\left(\frac{m_\alpha}{m_n} - 1\right)\left[3(5-3q) + \frac{m_\alpha}{m_n}(q-3)\right] {}_2F_1\left(\frac{1}{2}, \frac{1}{q-1}; \frac{5}{2} + \frac{1}{1-q}; \frac{m_\alpha}{m_n}\right)\right.$$
$$\left. + 2(3-2q)\left[(19-11q) + \frac{m_\alpha}{m_n}(1-q)\right] {}_2F_1\left(-\frac{1}{2}, \frac{1}{q-1}; \frac{5}{2} + \frac{1}{1-q}; \frac{m_\alpha}{m_n}\right)\right\}, \quad \text{(B.4)}$$

with
$$C_q = \frac{N_\alpha N_n \sigma_{12}^2 B_q}{(2-q)(3-2q)(9-5q)(11-7q)}\sqrt{\frac{8\pi k_B T}{m_\alpha}}.$$

Similarly, for $q > 1$, we have $v_{\alpha,\max} = v_{n,\max} = +\infty$. By using the integral transformation

$$\int_0^{+\infty} dv_2 F_n(v_2) v_2^2 \int_0^{v_2} dv_1 F_\alpha(v_1) v_1^2 \left(v_2 + \frac{v_1^2}{3v_2}\right) = \int_0^{+\infty} dv_1 F_\alpha(v_1) v_1^2 \int_{v_1}^{+\infty} dv_2 F_n(v_2) v_2^2 \left(v_2 + \frac{v_1^2}{3v_2}\right), \quad \text{(B.5)}$$

we have that

$$\nu_{q,\alpha n} = \frac{2 N_\alpha N_n m_\alpha^{\frac{3}{2}} m_n^{\frac{3}{2}} \sigma_{12}^2 B_q^2}{(k_B T)^3}\left\{\int_0^{+\infty} dv_2 F_n(v_2) \int_{v_2}^{+\infty} dv_1 \left(v_1^3 v_2^2 + \frac{v_1 v_2^4}{3}\right)\left[1 - \frac{(1-q)m_\alpha v_1^2}{2k_B T}\right]^{\frac{1}{1-q}}\right.$$
$$\left. + \int_0^{+\infty} dv_1 F_\alpha(v_1) \int_{v_1}^{+\infty} dv_2 \left(v_2^3 v_1^2 + \frac{v_2 v_1^4}{3}\right)\left[1 - \frac{(1-q)m_n v_2^2}{2k_B T}\right]^{\frac{1}{1-q}}\right\}$$

$$= \frac{4 N_\alpha N_n m_\alpha^{\frac{3}{2}} m_n^{\frac{3}{2}} \sigma_{12}^2 B_q^2}{(2-q)(3-2q)k_B T}\left\{\int_0^{+\infty} dv_2 \frac{v_2^2}{m_\alpha^2}\left[1 - \frac{(1-q)m_n v_2^2}{2k_B T}\right]^{\frac{1}{1-q}}\left[1 + \frac{(9-5q)m_\alpha v_2^2}{6k_B T}\right]\left[1 - \frac{(1-q)m_\alpha v_2^2}{2k_B T}\right]^{\frac{1}{1-q}+1}\right.$$
$$\left. + \int_0^{+\infty} dv_1 \frac{v_1^2}{m_n^2}\left[1 - \frac{(1-q)m_\alpha v_1^2}{2k_B T}\right]^{\frac{1}{1-q}}\left[1 + \frac{(9-5q)m_n v_1^2}{6k_B T}\right]\left[1 - \frac{(1-q)m_n v_1^2}{2k_B T}\right]^{\frac{1}{1-q}+1}\right\}$$

$$= \frac{4 N_\alpha N_n m_\alpha^{\frac{3}{2}} m_n^{\frac{3}{2}} \sigma_{12}^2 B_q^2}{(2-q)(3-2q)k_B T}\int_0^{+\infty} g_{\alpha n}(v_1) dv_1 \left\{\left[1 - \frac{(1-q)m_\alpha v_1^2}{2k_B T}\right]\left[1 - \frac{(1-q)m_n v_1^2}{2k_B T}\right]\right\}^{\frac{1}{1-q}}$$

$$= C_q \left\{\left(\frac{m_\alpha}{m_n}\right)^{\frac{1}{q-1}-\frac{3}{2}} A_{q,\alpha n} + (1-q)\left(\frac{m_\alpha}{m_n} - 1\right)\left[3(5-3q) + \frac{m_\alpha}{m_n}(q-3)\right] {}_2F_1\left(\frac{1}{2}, \frac{1}{q-1}; \frac{5}{2} + \frac{1}{1-q}; \frac{m_\alpha}{m_n}\right)\right.$$
$$\left. + 2(3-2q)\left[(19-11q) + \frac{m_\alpha}{m_n}(1-q)\right] {}_2F_1\left(-\frac{1}{2}, \frac{1}{q-1}; \frac{5}{2} + \frac{1}{1-q}; \frac{m_\alpha}{m_n}\right)\right\}, 1 < q < \frac{3}{2},$$
(B.6)

with

$$A_{q,\alpha n} = \frac{4\Gamma\left(\frac{1}{1-q} + \frac{7}{2}\right)\Gamma\left(\frac{2}{q-1} - \frac{3}{2}\right)}{3\sqrt{\pi}(7-5q)\Gamma\left(\frac{1}{q-1} - \frac{1}{2}\right)}\left\{2(7q-11)(q-3)(3q-5)\left(1 + \frac{m_\alpha}{m_n}\right) {}_2F_1\left(\frac{2}{q-1} - \frac{5}{2}, \frac{1}{q-1}; \frac{1}{q-1} - \frac{3}{2}; \frac{m_\alpha}{m_n}\right)\right.$$

$$+3(7q-11)(q-1)(5q-9)\left[1+\left(\frac{m_\alpha}{m_n}\right)^2\right]{}_2F_1\left(\frac{2}{q-1}-\frac{3}{2},\frac{1}{q-1};\frac{1}{q-1}-\frac{1}{2};\frac{m_\alpha}{m_n}\right)$$

$$-2(3q-5)(5q-7)(5q-9){}_2F_1\left(\frac{2}{q-1}-\frac{7}{2},\frac{1}{q-1};\frac{1}{q-1}-\frac{5}{2};\frac{m_\alpha}{m_n}\right)\Bigg\}. \tag{B.7}$$

**Appendix C**

Eq.(31) is calculated as follows.

$$\mathbf{F}_{ei} = -\frac{N_e N_i Z_i^2 e^4 \ln\Lambda}{4\pi\varepsilon_0^2 m_e} B_{q,e} \left(\frac{m_e}{2\pi k_B T_e}\right)^{\frac{3}{2}} \int d\mathbf{v}\,\frac{\mathbf{v}}{v^3}\left\{1+\frac{m_e \mathbf{v}\cdot\mathbf{u}}{k_B T_e}\left[1-(1-q)\frac{m_e \mathbf{v}^2}{2k_B T_e}\right]^{-1}\right\}\left[1-(1-q)\frac{m_e \mathbf{v}^2}{2k_B T_e}\right]^{\frac{1}{1-q}}$$

$$= -\frac{N_e N_i Z_i^2 e^4 B_{q,e} \ln\Lambda}{4\pi\varepsilon_0^2 k_B T_e}\left(\frac{m_e}{2\pi k_B T_e}\right)^{\frac{3}{2}} \int d\mathbf{v}\,\frac{\mathbf{v}\mathbf{v}\cdot\mathbf{u}}{v^3}\left[1-(1-q)\frac{m_e \mathbf{v}^2}{2k_B T_e}\right]^{\frac{1}{1-q}-1}. \tag{C.1}$$

Because $\mathbf{u}=(0, 0, u_z)$, we obtain that

$$F_{ei,z} = -\frac{N_e N_i Z_i^2 e^4 B_{q,e} \ln\Lambda}{4\pi\varepsilon_0^2 k_B T_e}\left(\frac{m_e}{2\pi k_B T_e}\right)^{\frac{3}{2}} u_z \int_{-v_{max}}^{v_{max}} \frac{v_z^2}{v^3}\left[1-(1-q)\frac{m_e \mathbf{v}^2}{2k_B T_e}\right]^{\frac{1}{1-q}-1} d\mathbf{v}$$

$$= -\frac{N_e N_i Z_i^2 e^4 B_{q,e} \ln\Lambda}{3\varepsilon_0^2 k_B T_e}\left(\frac{m_e}{2\pi k_B T_e}\right)^{\frac{3}{2}} u_z \int_0^{v_{max}} \left[1-(1-q)\frac{m_e \mathbf{v}^2}{2k_B T_e}\right]^{\frac{1}{1-q}-1} v\,dv. \tag{C.2}$$

For $0 < q < 1$, we have $v_{max}=[2k_B T_e/(1-q)m_e]^{1/2}$, and so,

$$F_{ei,z} = -\frac{\sqrt{2}m_e^{\frac{1}{2}} N_e N_i Z_i^2 e^4 u_z \ln\Lambda}{12\varepsilon_0^2 (\pi k_B T_e)^{\frac{3}{2}}}(1-q)^{\frac{1}{2}}(3-q)(5-3q)\frac{\Gamma\left(\frac{1}{1-q}+\frac{1}{2}\right)}{4\Gamma\left(\frac{1}{1-q}\right)}. \tag{C.3}$$

For $1 < q < 5/3$, we have $v_{max}=+\infty$, and so

$$F_{ei,z} = -\frac{N_e N_i Z_i^2 e^4 B_{q,e} \ln\Lambda}{3\varepsilon_0^2 k_B T_e}\left(\frac{m_e}{2\pi k_B T_e}\right)^{\frac{3}{2}} u_z \int_0^{+\infty} \left[1-(1-q)\frac{m_e \mathbf{v}^2}{2k_B T_e}\right]^{\frac{1}{1-q}-1} v\,dv$$

$$= -\frac{\sqrt{2}m_e^{\frac{1}{2}} N_e N_i Z_i^2 e^4 u_z \ln\Lambda}{12\varepsilon_0^2 (\pi k_B T_e)^{\frac{3}{2}}}\frac{(q-1)^{\frac{3}{2}}\Gamma\left(\frac{1}{q-1}\right)}{\Gamma\left(\frac{1}{q-1}-\frac{3}{2}\right)},\ 1<q<\frac{5}{3}. \tag{C.4}$$